\documentclass[preprint]{aastex6}
\RequirePackage{natbib}
\newcommand{\mbf}[1]{\ensuremath{\mbox{\boldmath{$#1$}}}}

\usepackage{graphicx}
\def\bul{\raise.2ex\hbox{$\bullet$}}
\def\solar{\odot}

\renewcommand{\aap}{    {\it Astron. Astrophys.}}

\renewcommand{\apjs}{   {\it Astrophys. J. Suppl. Ser.}}
\chardef\us=`\
\bibpunct[; ]{(}{)}{;}{a}{,}{;}
\begin{document}
\title{Generation of a North/South Magnetic Field Component from Variations in the Photospheric Magnetic Field}
\author{Roger K. Ulrich and Tham Tran}
\affil{Department of Physics and Astronomy, University of California, Los Angeles CA 90095}
\email{ulrich@astro.ucla.edu}
\begin{abstract}
We address the problem of calculating the transverse magnetic field in the solar wind outside of the hypothetical sphere called the source surface where the solar wind originates.  This calculation must overcome a widely used fundamental assumption about the source surface -- the field is normally required to purely radial at the source surface.   
Our model rests on the fact that a change in the radial field strength at the source surface is a change in the field line density.  Surrounding field lines must move laterally in order to accommodate this field line density change.  As the outward wind velocity drags field lines past the source surface this lateral component of motion produces a tilt implying there is a transverse component to the field. 
  An analytic method of calculating the lateral translation speed of the field lines is developed.  We apply the technique to an interval of approximately two Carrington rotations at the beginning of 2011 using 2-h averages of data from the {\it Helioseismic Magnetic Imager} instrument on the {\it Solar Dynamics Observatory} spacecraft.  We find that the value of the transverse magnetic field is dominated on a global scale by the effects of high latitude concentrations of field lines being buffetted by supergranular motions.  
\end{abstract}
\section{Introduction}
The strength of terrestrial geomagnetic storms depends on the value of the solar wind magnetic field near Earth in the southward direction \citep{1999SSRv...88..529G} called $B_{\rm s}$.  This quantity is usually used to denote an intense, long-duration, and southward magnetic field.  In interplanetary space the associated quantity is $B_{\rm z}$ which is the momentary value of the magnetic field in a direction perpendicular to the ecliptic.  The origin of $B_{\rm z}$ is not well understood since the observed values correlate in a weak and complex manner with specific solar quantities such as the sunspot number and sector boundaries \citep{2014JGRA..119..658Z}.  While stronger excursions in $B_{\rm s}$ and $B_{\rm z}$ are associated with solar events such as coronal mass ejections, these authors suggested that many of the excursions and discontinuities in $B_{\rm z}$ come from Alfv\'enic turbulence generated by the Sun and evolved between the Sun and earth. 

At the point in the heliosphere where the solar wind originates and the gas component of the plasma begins to dominate the magnetic field it is usually assumed that $B_\theta$ (the component in the north/south (N/S) direction at the Sun) vanishes.  Although the quantities needed at Earth are $B_{\rm z}$ or $B_{\rm s}$, the extension from the Sun to Earth by way of the solar wind is complicated and we focus on the solar quantity $B_\theta$.  The transition from the relatively static corona to the solar wind occurs at the source surface and while there are a number of ways to model the coronal zone 
\citep{1969SoPh....9..131A,1971CosEl...2..232S,1994SoPh..151...91Z}, all these methods include the assumption that the magnetic field is radial at the source surface.  The first of the above models is called the potential-field source-surface (PFSS) model and assumes there are no electric currents in the volume between the photosphere and the source surface. The second and third models include an intermediate surface called either the `source surface' (the quotes indicate that this is not the actual source surface) or cusp surface on which there are electric currents. The third of the three models also includes volume currents between the photosphere and the cusp surface following the model of \citet{1986ApJ...306..271B}. The recent work by \citet{2015ApJ...803L...1J} bypasses the assumption of the vanishing transverse field at the source surface by assuming that the transverse fields from the cusp surface persist into the interplanetary region of the heliosphere.  This model is discussed briefly below.  Another approach was taken by \citet{1978SoPh...60...83S} who took the source surface to be non-spherical and be located on a surface where the field strength is a constant whose value can be adjusted.  Unfortunately, time variability in the location of the source surface cannot be treated with the method we develop.  We describe in the current paper an alternate cause for a non-zero $B_\theta$ which comes from the time dependence of $B_r$ and the radial motions of the solar wind.

We start by modifying the assumption that the heliospheric magnetic field is purely radial at the source surface \citep{1969SoPh....9..131A,1969SoPh....6..442S}.  In particular, we note that temporal variation $B_r$ at the source surface due to changes in the underlying photospheric magnetic configuration causes the field lines passing through the source surface to drift in a transverse direction in order to accommodate an increase or decrease in the magnetic flux within any given area.  The generation of transverse field from field line drift has been proposed and discussed by \citet{1989GeoRL..16....1J} in the context of field line dragging due to supergranulation velocities.
Our model is similar in that the deduced transverse field is given by the product of the radial field and the ratio of the transverse velocity to the radial velocity as was assumed by those authors.  The difference for our model is that we derive the transverse velocity of the magnetic field from the changes in the field strength over the solar surface due to the emergence or subsidence of the photospheric field lines that penetrate out to the source surface and force a lateral displacement of the field lines at that position.  

\begin{figure}
\begin{center}
\hspace*{-0.0in}\parbox{6.5in}{\hspace{0.0in}
\resizebox{6.5in}{!}{\includegraphics{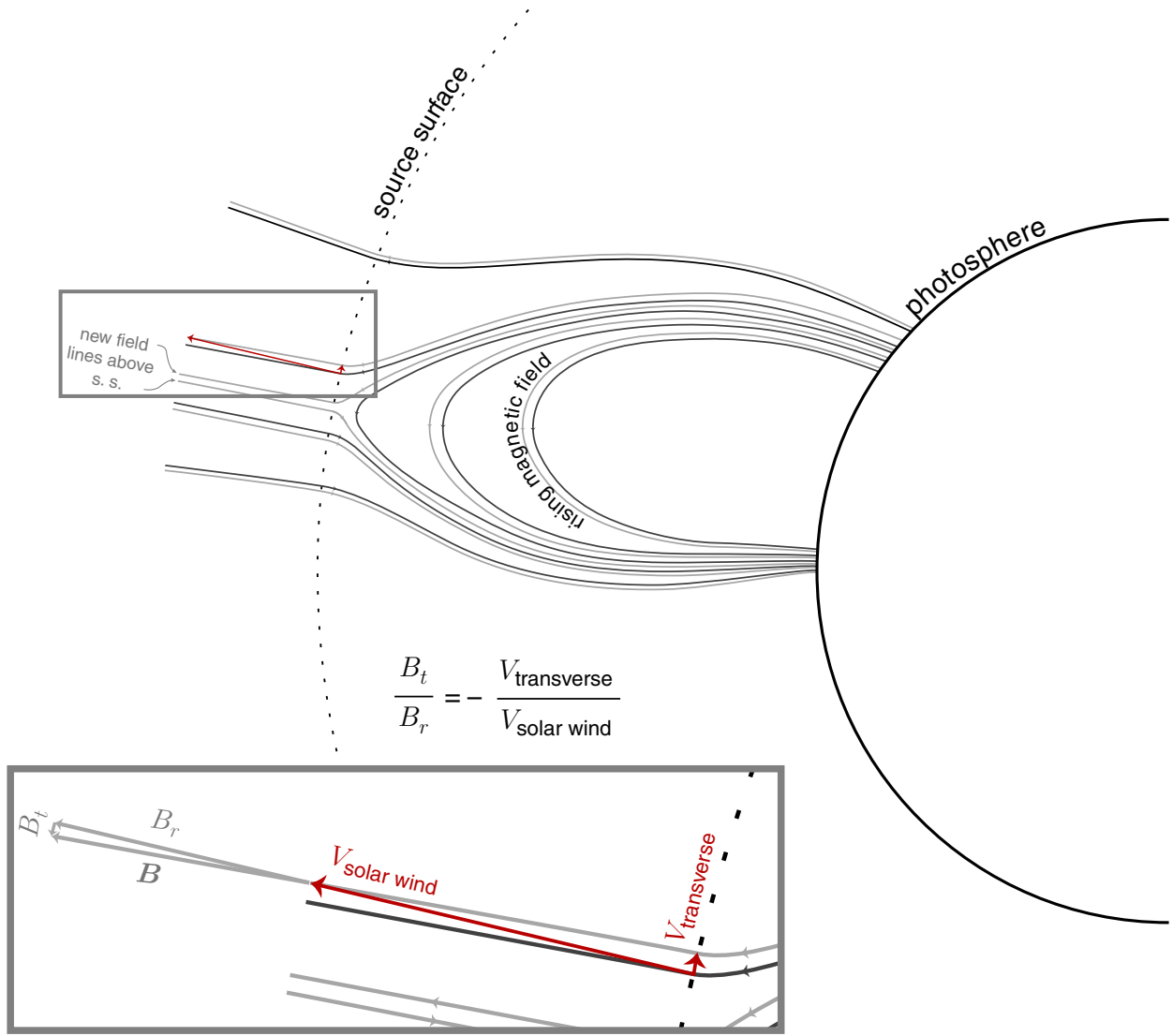}}
\caption{\baselineskip=15pt
This sketch of a generic magnetic geometry for two successive times
shows how a rising magnetic loop can push field lines out of the 
potential field region through the source surface causing the displacement
of adjacent field lines that have already extended beyond the source surface (s.s.).
The darker lines represent a configuration at the earlier time and the grey
field lines represent the subsequent configuration.  The pair of lines labeled
``new field lines above s.s.'' illustrate the part of the loop which has been
drawn out into the solar wind region.  The red arrows show the radial direction
of the solar wind and the transverse motion of the field lines caused by the
newly emergent field.  The grey box gives a magnified view of the affected field lines with the resulting radial and transverse components.}
\label{figzero}
}
\end{center}
\end{figure}

We give an example of a generic geometry illustrating the process of transverse
field generation from temporal changes in the potential field structure. Figure~\ref{figzero} shows the expansion of a magnetic loop that inserts new
field lines into the solar wind zone of the heliosphere.  As the loop reaches
and passes through the source surface, the top of the loop is stretched radially
outward generating a pair of inward and outward pointed field lines.  To accomdate 
this new field, the adjacent field lines must move in a transverse direction.  The
velocity of this transverse motion depends on the rate at which the field penetrates the source surface, {\it i.e.}\ it depends on the time derivative of $B_r$ at the source surface.  The simplified sketch of Figure~\ref{figzero} omits 
many complicating effects such as differential rotation, sunspot evolution, and convective motions in the photosphere.  The algorithm developed below can
include such processes as long as they are part of the photospheric magnetogram
record.

After a transverse field component has been generated this way the solar wind drags these field lines with their non-radial components out into the heliosphere.  The phase relationship between the transverse field and the transverse velocity is that of Alfv\'en waves which can propagate relative to
the solar wind plasma.  In addition, when these waves encounter a corotating 
interaction region where a more rapidly moving stream meets an overlying slower moving stream the transverse field can be compressed and strengthened or otherwise changed in a manner similar to the turbulence model discussed by \citet{2001JGR...10615929H} in the context of field strength variability observed by {\it Ulysses}.  Treatment of such effects will require model development that is
beyond the scope of the present work.

A recent publication by \citet{2015ApJ...803L...1J} has derived a similar quantity $B_n$ using the current-sheet source surface (CSSS) model of \citet{1995JGR...100...19Z}.  Although the CSSS model like the PFSS model assumes $B_\theta$ and $B_n$ are zero at the source surface, \citet{2015ApJ...803L...1J} adopt the value from the inner cusp surface which they denote as the flux release surface as representative of what escapes into the heliosphere as part of the solar wind.  When they compare the result to observations, they adjust the value by an arbitrary factor of 0.02 to obtain a slope of one in a correlation diagram.  
Their data are on a cadence of less than once per day and the observed field strengths at 1 AU have also been smoothed to a time scale of one day. 
We do not use the CSSS approach because it imposes a non-linearity in the relationship between magnetic field values at the solar surface and those at the source surface due to the fact that the field direction is reversed at the cusp surface and again at the source surface for one of the two field directions.  The consistency between this double reversal of the field direction and the framework used below of field line motion developed by \citet{1966SSRv....6..147S} is not established.  While this approach is helpful in reproducing some aspects of the field structure between these two surfaces it also places the source surface at a much greater distance from the photosphere which would make our use below of a 2-h cadence suspect due to the large propagation times.  We find non-zero but much smaller values of $B_\theta$ compared with \citet{2015ApJ...803L...1J}.

As indicated above, our theory of the transverse component depends on the rate of change of $B_r$ at the source surface.  The data from the {\it Helioseismic and Magnetic Imager}  \citep[{\it HMI}; ][]{2012SoPh..279..295L,2014SoPh..289.3483H} instrument on the {\it Solar Dynamics Observatory} (SDO) covers the whole visible surface at a high temporal cadence and is obtained with few temporal gaps.  These qualities allow us to study time derivatives of the magnetic field in a way that has never before been possible.  This dataset shows that the global solar magnetic field is in a state of continuous variation on a time scale that is shorter than one day.  The resulting transverse fields vary in a way that is not easily related to the major features such as sunspots, active regions, or plages.

\section{Transverse Magnetic Field Generation by Field Line Displacement}\label{sectwo}
The source surface is 
where open magnetic field lines are assumed to be drawn outward by the solar wind so that they exit into the heliosphere.
Evolution of the solar surface magnetic field causes continuous change in the field at the source surface so
that the frozen-in field lines are dragged laterally at the same time as they are being pulled outward.  The
concepts of lateral dragging of magnetic field lines were developed by \citet{1966SSRv....6..147S}.  The idea that the transverse motion of the field lines could yield an estimate for the strength of the transverse field was proposed by \citet{1968PhRvL..21...44J} who assume that the ratio of transverse field to radial field is the same as the ratio of transverse velocity to radial solar wind velocity. 
Using this idea we can find the transverse field just outside the source surface by calculating the horizontal velocity of
the field lines and then apply the Jokipii-Parker assumption.  A related idea was developed by \citet{2007ApJ...663..583G} who discussed the lateral displacement of the field lines but imposed a condition that there is no horizontal shear -- a condition which explicitly rules out the Jokipii-Parker mechanism.

We wish to find the horizontal velocity
\begin{eqnarray}
\mbf{v}_{\rm h}&=&v_\theta \mbf{a}_\theta+v_\phi\mbf{a}_\phi
\end{eqnarray}
where $\mbf{a}_\theta$ and $\mbf{a}_\phi$ are unit vectors in the $\theta$ and $\phi$ directions.
We start from the frozen-in equation for the field in a perfectly conducting fluid
\begin{eqnarray}
{\partial B_r \over \partial t} = [\nabla \times (\mbf{v}\times\mbf{B})]_r 
\end{eqnarray}
and expand the vectors assuming only $B_r$ is non-zero,
\begin{eqnarray}
[\nabla \times (\mbf{v}\times\mbf{B})]_r &=& {1\over r \sin \theta}\left[ {\partial\over\partial\theta}(\sin\theta (\mbf{v}\times\mbf{B})_\phi) - {\partial\over\partial\phi}(\mbf{v}\times\mbf{B})_\theta\right]\ .\\
\noalign{\noindent We then expand the cross products as}
\sin\theta (\mbf{v}\times\mbf{B})_\phi &=&  - \sin\theta \;v_\theta\;B_r \\
\noalign{\noindent and}
(\mbf{v}\times\mbf{B})_\theta &=& v_\phi\;B_r
\end{eqnarray}
so that the frozen field formula becomes
\begin{eqnarray}
{\partial B_r \over \partial t} &=& [\nabla \times (\mbf{v}\times\mbf{B})]_r = - {1\over r \sin \theta}\left[ {\partial\over\partial\theta}( \sin\theta \;v_\theta\;B_r ) + {\partial\over\partial\phi} (v_\phi\;B_r)  \right]\ .
\label{dBrdtequn}
\end{eqnarray}

\bigskip
\noindent In Equation~(\ref{dBrdtequn}) we will calculate $\dot B_r = \partial B_r /\partial t$ from time series of magnetic models connecting the photosphere to the spherical source surface. These models will also provide $B_r$.  To find $v_\theta$ and $v_\phi$ we solve Equation~(\ref{dBrdtequn}) using a horizontal potential function $\eta$ which determines the product of $B_r$ and the horizontal velocity $\mbf{v}_{\rm h}$.  The desired potential function is then defined by
\begin{eqnarray}
B_r\mbf{v}_{\rm h} &=& \mbf{\nabla}_{\rm h}\eta
\end{eqnarray}
where the horizontal gradient is 
\begin{eqnarray}
\mbf{\nabla}_{\rm h} &=& {\mbf{a}_\theta\over r}{\partial\over\partial\theta}
+{\mbf{a}_\phi\over r\sin\theta}{\partial \over \partial\phi}\ .
\end{eqnarray}
The horizontal velocities are then given by
\begin{eqnarray}
v_\theta\,B_r &=& r^{-1}\partial \eta/\partial \theta\\
\noalign{\noindent and}
v_\phi\,B_r &=& (r\sin\theta)^{-1}\partial\eta/\partial \phi\ .
\end{eqnarray}
Inserting these into Equation~(\ref{dBrdtequn}) we get
\begin{eqnarray}
{\partial B_r \over \partial t} &=&  - {1\over r \sin \theta}\left[ {\partial\over\partial\theta}\left( \sin\theta \;{1\over r}{\partial \eta\over\partial \theta }\right) + {\partial\over\partial\phi}\left({1\over r\sin\theta}{\partial\eta\over\partial\phi}\right)  \right]\ ,
\end{eqnarray}
and using the angular part of the Laplacian expressed in spherical coordinates 
\begin{eqnarray}
\nabla^2_{\rm h} &=& {1\over r^2\sin\theta}{\partial\over\partial\theta}\left(\raise 0.8ex\hbox{$\sin\theta$}{\partial\over\partial\theta}\right)
+{1\over r^2\sin^2\theta}{\partial^2\over\partial\phi^2}
\end{eqnarray}
we obtain
\begin{eqnarray}
{\partial B_r \over \partial t} &=& - \nabla^2_{\rm h} \eta \ .
\label{etaequation}
\end{eqnarray}
Since $\partial B_r/\partial t$ can be calculated from the potential field source surface solutions, this equation can be solved using standard techniques based on spherical harmonics.  Then
\begin{eqnarray}
v_\theta = {1\over r B_r}{\partial \eta\over\partial\theta}\quad &,& \quad
v_\phi = {1\over r B_r\; \sin\theta}{\partial\eta\over\partial\phi}
\end{eqnarray}
and with the assumption that ${B_\theta/ B_r}=-{v_\theta / v_{\rm wind}}$ 
and ${B_\phi/ B_r}=-{v_\phi / v_{\rm wind}}$ we get
\begin{eqnarray}
B_\theta = -{1\over r v_{\rm wind}}{\partial\eta\over\partial\theta}\quad &,&
\quad B_\phi = -{1\over r \sin\theta\; v_{\rm wind}}{\partial\eta\over\partial\phi}\ .
\label{Bthetaeqn}
\end{eqnarray}
Note that $\eta$ has dimensions of magnetic field times distance times velocity.

\bigskip

\section{Spherical Harmonic Solution for the Transverse Magnetic Field}
We use a definition of the spherical harmonics $Y^m_\ell(\theta,\phi)$ which constitute
an orthonormal set of basis functions.  The definition we use is
\begin{eqnarray}
 Y^m_\ell(\theta,\phi)&=& N^m_\ell\;P^m_\ell(\cos\theta)\exp(i m \phi)\\
\noalign{\noindent with}
N^m_\ell &\equiv& (-1)^{(m+|m|)/2}\left({2\ell+1\over4\pi}{(\ell-|m|)!\over(l+|m|)!}\right)^{1/2}\ .
\end{eqnarray}
The spherical harmonics have the following orthonormality properties:
\begin{eqnarray}
\int_0^{2\pi}d\phi\int_0^\pi\,d\theta\,\sin\theta\,Y_\ell^m(\theta,\phi)\,\overline{Y_{\ell^\prime m^\prime}(\theta,\phi)}
&=& \delta_{\ell,\ell^\prime}\,\delta_{m,m^\prime}\ .
\end{eqnarray}
We expand $\eta(\theta,\phi)$ in spherical harmonices with coefficients $E_{\ell,m}$:
\begin{eqnarray}
\eta &=& \sum_{\ell m} E_{\ell,m}Y_{\ell,m}(\theta,\phi)\ .
\end{eqnarray}
The quantity that drives the solution is the time derivative of the radial component of the magnetic field at the source surface $\dot B_{r\,{\rm ss}}\equiv (\partial B_r/\partial t)_{\rm ss}$ which we calculate numerically from first 
differences in the time series.  Inserting this expansion into Equation (\ref{etaequation}) and using the orthonormality conditions we obtain the $E_{\ell,m}$ coefficients
as 
\begin{eqnarray}
E_{\ell,m}&=&{r^2\over \ell(\ell+1)}\int_{2\pi} d\phi\;\int_0^\pi \dot B_{r\,ss}\,\overline{Y_\ell^m(\theta,\phi)}\sin\theta\,d\theta\equiv {r^2\over \ell(\ell+1)}\;C_{\ell m}\ .
\end{eqnarray}

\section{Observations and Reduction Methods}
\label{methods}
An initial study of the photospheric fields as a source for the transverse components of $\mbf{B}$ was done using data from Mt.\ Wilson Observatory \citep{2002ApJS..139..259U} and the {\it Michelson Doppler Imager} \citep[{\it MDI}; ][]{1995SoPh..162..129S} on the {\it Solar and Heliospheric Observatory} (SoHO).  The observations from these two systems did not permit time resolution shorter than about one day and this cadence was inadequate for our study.  We found there to be substantial variations at the basic resolution limit of one day so that we could not be confident that the observed changes were of solar origin, and if they were of solar origin we felt we could not properly determine the nature of the changes without better temporal resolution.  The high cadence of the magnetic fields measured by the {\it Helioseismic and Magnetic Imager} \citep[{\it HMI}; ][]{2012SoPh..279..295L,2014SoPh..289.3483H} system on the {\it Solar Dynamics Observatory} permits us to reach the one-day time scale and shorter with confidence.

We started with sequences of 4098$\times$4098 hmi.M\_720s $B_{\rm los}$ images and immediately rebinned these to 1024$\times$1024 images.  Although these fields do not agree with other field measurements such as from the MDI system on SoHO \citep{2012SoPh..279..295L}, we have not applied any adjustment.    We then divided the time series into 2-h windows and shifted each image in the window to the central time correcting for the effect of differential rotation.  The images are then averaged together and downsized these to a 512$\times$512 array.  This set of operations is called a derot mean.  For the present study we began by using a 6-h window but ended up using a 2-h windowed when it became apparent that the 6-h series is temporally undersampled. We refer to the 6-h series only briefly below. 

According to \citet{2005A&A...435.1123W} a typical Alfv\'en speed in the potential field portion of the corona is about 600 km s$^{-1}$ so that a disturbance could propagate a distance of 1.5$R_\solar$ in less than 30 min.  Thus for our application using the PFSS model with a source surface at 2.5$R_\solar$, the  Alfv\'en wave travel time across the region interior to the source surface is comparable to or shorter than our 2-h time step. Because we are applying the transverse field algorithm at the source surface, which is at the base of the solar wind, the finer structure from the original HMI images decays in a potential field model to a small enough level that use of 512$\times$512 and 256$\times$256 grids instead of the original 4096$\times$4096 grid has little impact on the result at the source surface.  In fact estimates of the solar wind speed such as that by \citet{1990ApJ...355..726W} depend on empirical parameters calculated from photospheric magnetograms having spatial resolution comparable to 256$\times$256.  Use of higher resolution photospheric maps would require recalibration of these empirical parameters.

Our goal is to estimate the time rate of change of the radial magnetic field component on the source surface.  We have applied the potential-field source-surface method of \citet{1969SoPh....9..131A} to this task.  We have adopted this method due to its simplicity and linearity.  The linearity allows us to start with time derivatives of the photospheric field and calculate the corresponding time derivatives at the source surface from the same potential field model as is applied to the fields themselves.  This virtue allows us to compare two methods: the expansion of photospheric differences (we denote as the PD method) and the direct difference of successive expansions (we denote this as the direct difference or DD method).  We can compare how numerical and other errors in the photospheric fields impact the source surface time derivative when each of these methods is used.  This comparison shows that there are differences in the results which range from moderate to rather large and we concluded that the extrema produced by the PD method were larger than from the DD method so that we have adopted the DD approach as preferred.  However, we have retained the option of using the PD method in case further investigation shows that it gives better results when incorporated into a more complete analysis.

The method developed by \citet[hereinafter called the UB method]{2006SoPh...235...17U} includes an indication of time dependence over a time scale of one Carrington rotation (27.2753 days).  However, the solar surface changes much more quickly than this so the snapshot maps created with this method miss much of the variation that occurs.  For those portions of the solar surface that are unseen, there is no approach that can show more rapid evolution than this.  The use of a flux transport model is a form of extrapolation and cannot show the actual state of the solar surface.  However, we do have real time observations from HMI that can be included.  We do utilize the UB method to resolve the $\mbf{B}$ vector into its $\mbf{r}$ and $\mbf{\phi}$ components.  We then remove an average $B_\phi$ component from the observed $B_{\rm los}$ value so that we can include a best estimate of the $B_r$ component at each pixel.  We call the result a {\sf Bro} image because these are our best estimates of the observed value of $B_r$.  We use an interpolation method to insert the {\sf Bro} valuations into the snapshot map and refer to the result as a hybrid map.

The sequence we have developed has a 2-h time interval and includes higher frequencies than we initially wished to study in this effort.   To mitigate the highest frequency uncertainties such as could come from registration and interpolation in the mapping process we have applied a temporal filter using a weighting of the five snapshot maps nearest to each final map.  Taking the current time to be $t_i$, the weights at $(t_{i-2}, t_{i-1}, t_i, t_{i+1}, t_{i+2})$ are (0.075,\,0.175,\,0.5,\,0.175,\,0.075) where the time interval between each of the times is 2 h.  During periods for the current series there are two gaps in the data.  During these gaps we have calculated the differences from the nearest pair of observations offset symmetrically from the map time for the photospheric time differences.  In the approach where we use the direct differences of potential field expansions we find maps at the missing times by averaging the two symmetrically placed maps nearest the missing time then carry out the direct difference on the derived PFSS solutions as before.  For these gap periods the amplitude of $\dot B_r$ variation is reduced.
We also include a reduction in which the temporal smoothing was not done.  This case allows us to study the temporal power spectrum in an effort to identify the cause or causes of the variability present in the time series.

\begin{figure}
\begin{center}
\hspace*{0.1in}\parbox{6.0in}{\hspace{1.0in}
\resizebox{4.2in}{!}{\includegraphics{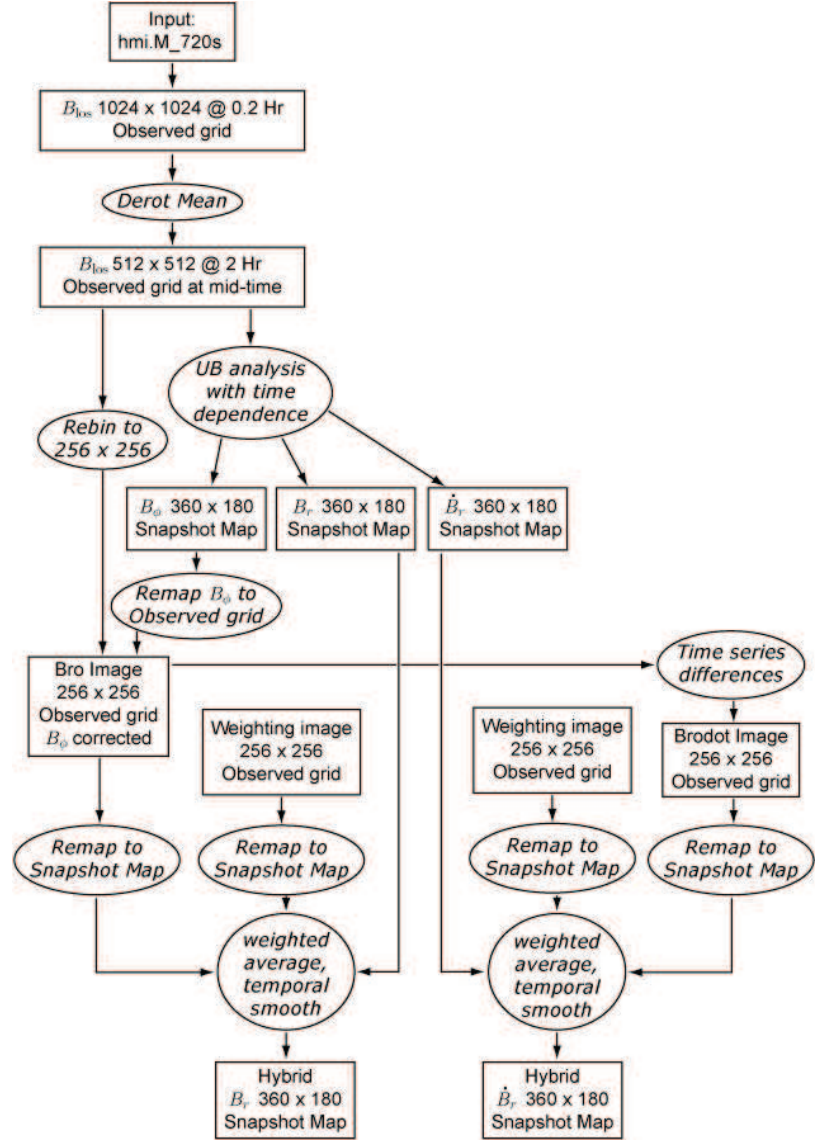}}
\caption{\baselineskip=15pt
This chart shows the flow of the data through our processing system.  Data
products at various stages are shown in rectangular boxes, processing steps
are shown in the ellipses with a slanted font.  Two data formats are used: 
(1) an image format that represents the magnetic field on an observed grid which is on the plane of the sky with a minor adjustment due to the differential rotation correction and (2) a snapshot map format that represents the data as a function of heliographic coordinates of latitude and longitude.  The zero point longitude of the snapshot map is taken as the Carrington longitude at the center of the solar disk at the mapping time.  All other longitudes are calculated from the geometric offset and are the central Carrington longitude plus the central meridian angle.  These are not Carrington longitudes except for the central longitude.}
\label{figzeroa}
}
\end{center}
\end{figure}

The reduction sequence is summarized in the flow chart given in Figure~\ref{figzeroa}. Specific steps are explained here:
\begin{enumerate}
\item The 512$\times$512 2-h averaged images are rebinned to 256$\times$256 and combined with UB analysis using time dependent terms.  The observed frame images are combined into snapshot maps: $B_r$, $B_\phi$, $\dot B_r=dB_r/dt$, and $\dot B_\phi = dB_\phi/dt$ (called 2CRtd maps).  For occasional latitude, longitude combinations the UB analysis fails and we use results from a backup analysis without the time dependent terms (called 1CR maps). Those cases do not provide $\dot B_r$.
\item The $B_\phi$ snapshot maps are transformed back to observed grid images where they are combined with the processed $B_{\rm los}$ images and are projection corrected to give {\sf Bro}, the radial component $(B_r)_{\rm obs}$.
\item {\sf Bro} images are used to calculate {\sf Brodot} images by numerical time differencing on a 2-h basis.
\item {\sf Bro} and {\sf Brodot} images are combined with $B_r$ and $dB_r/dt$ snapshot maps to yield hybrid\_{\sf Br} and hybrid\_{\sf Brdot} snapshot maps.
\end{enumerate}

The merging of the {\sf Bro} and {\sf Brodot} images into the corresponding snapshot maps uses a weighting map that is defined on the observed grid then transformed to the snapshot format.
The weighting function is designed to provide a smooth transition between the slowly varying snapshot maps and the rapidly varying observed fields.  However, the snapshot maps include information from the advanced time image so that they have a good representation of the average fields in strong field regions prior to their appearance on the observed grid.  This avoids the unbalanced influence of the appearance of the leading region of a bipolar group.  We also taper the
weight toward the limb so that the new regions do not influence the hybrid map
until the trailing components can be seen.
The weight function we use depends on both the center-to-limb angle $\rho$ and and on the central meridian angle CMA:
\begin{enumerate}
\item On the observed grid for which we have values of {\sf Bro} we create a weight image $w_1$ whose value is 1.0 for $\rho<65^\circ$ and goes linearly with $\rho$ to 0.0 at $\rho=90^\circ$.
\item The weight image is transformed to the snapshot grid with the same algorithm applied to {\sf Bro}.
\item On the snapshot map we limit {\sf Bro} data to points where $|{\rm CMA}|<75^\circ$.  For each latitude starting at the point on the snapshot map where $|{\rm CMA}|=75^\circ$ we calculate a second weight $w_2$ whose value is 0.0 and whose value is increased by a regular increment chosen to be about 0.03 for each snapshot grid point going to lower $|{\rm CMA}|$.
\item The weight applied to the {\sf Bro} data is the lesser of $w_1$ and $w_2$.
\end{enumerate}
This merging algorithm was designed to avoid the excessive noise near the limb while retaining as much information as possible about the polar regions while at the same time allowing a larger buffer nearer the equator where active regions are important. 

\section{A Sample Reduction}

\begin{figure}
\begin{center}
\hspace*{-0.5in}\parbox{7.5in}{\hspace{0.5in}
\resizebox{7.5in}{!}{\includegraphics{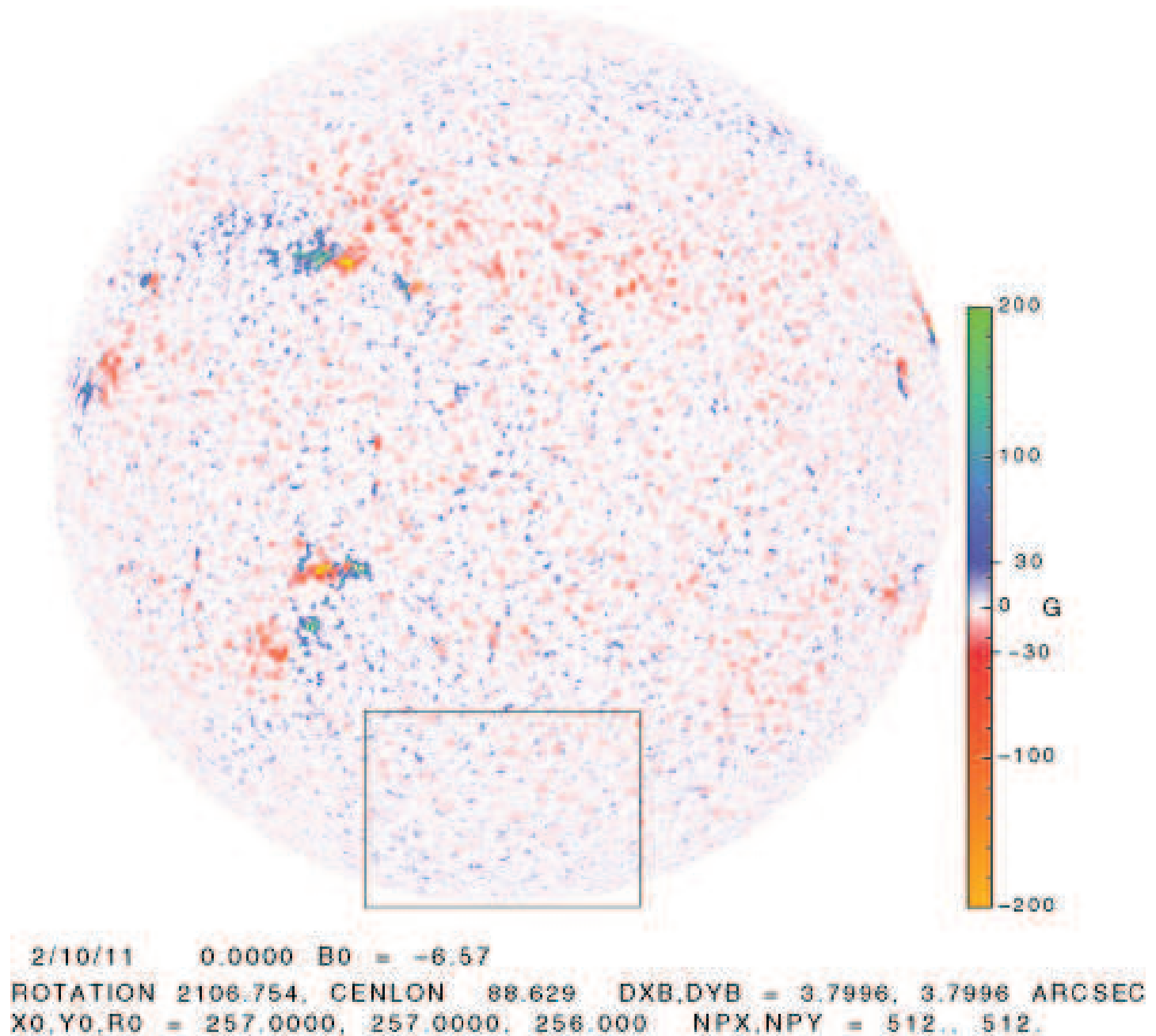}}
\caption{\baselineskip=15pt
This figure shows a sample 512$\times$512, 2-h averaged frame from the HMI system showing the line-of-sight component of the magnetic field.  The scale for this image
has been set so that full red and blue are at 30 gauss (G).  Higher fields typical of active regions use the orange and green color codes but fields that strong are not found in the polar areas.  The associated video ({\tt hmi512SouthPole.mp4}) is extracted from
the square at the bottom and uses a red/blue encoding where full red and blue are $\pm20$ G.  The rapid variation in
the pattern of field is a manifestation of the photospheric field process called the ``magnetic carpet" \citep{1998ASPC..154..345T} which is most often studied nearer the Sun's equator.  The video is 
at: 
{\tt http://www.astro.ucla.edu/\~{}ulrich/btss/
hmi512SouthPole.mp4}
.  The frame rate is 6 s$^{-1}$ with each frame representing 2 h of elapsed time.  The full sequence of 556 images covers a period of 46 days.  The supplementary data images have
a strongly yellow band across the bottom that comes from pixels off the limb.
}
\label{figone}
}
\end{center}
\end{figure}

We have selected a sample time period starting 7 January 2011 and ending 2 March 2011 to apply our approach.  It is difficult to illustrate the data in its various stages of reduction using static snapshots so we include several videos.  As a 
preliminary we examine the photospheric magnetic field, especially at high latitudes.  These turn out to be a dominant
factor in producing short term variability in the Sun's dipole field strength and orientation.  For the sample time
period, the south pole regions are visible while the north pole regions need to be estimated from our filling methods.  We
illustrate a sample 2-h average line-of-sight HMI magnetogram in Figure~\ref{figone}.  The color code has been chosen to
make the low magnetic field values evident at the expense of saturating the high field in active regions.  
The scale of the color encoding is not far from the noise levels of the HMI magnetograms and some details of the images are influenced by this noise. 
The rectangle at bottom expands the region around the south pole.  The supplementary data video
{\tt hmi512SouthPole.mp4} shows that the magnetic field has the type of structure and variability seen in the equatorial regions and which has been described as a magnetic carpet by \citet{1998ASPC..154..345T}.  Note that the fields illustrated in this figure have not been corrected for the fact that they are nearly radial and have a large multiplicative factor due their proximity to the solar limb.  Convective effects near the solar surface cause these concentrated field regions to vary on the time scale of the supergranulation which is less than a day. 

\begin{figure}\begin{center}
\hspace*{-0.5in}\parbox{7.5in}{
\resizebox{7.5in}{!}{\includegraphics{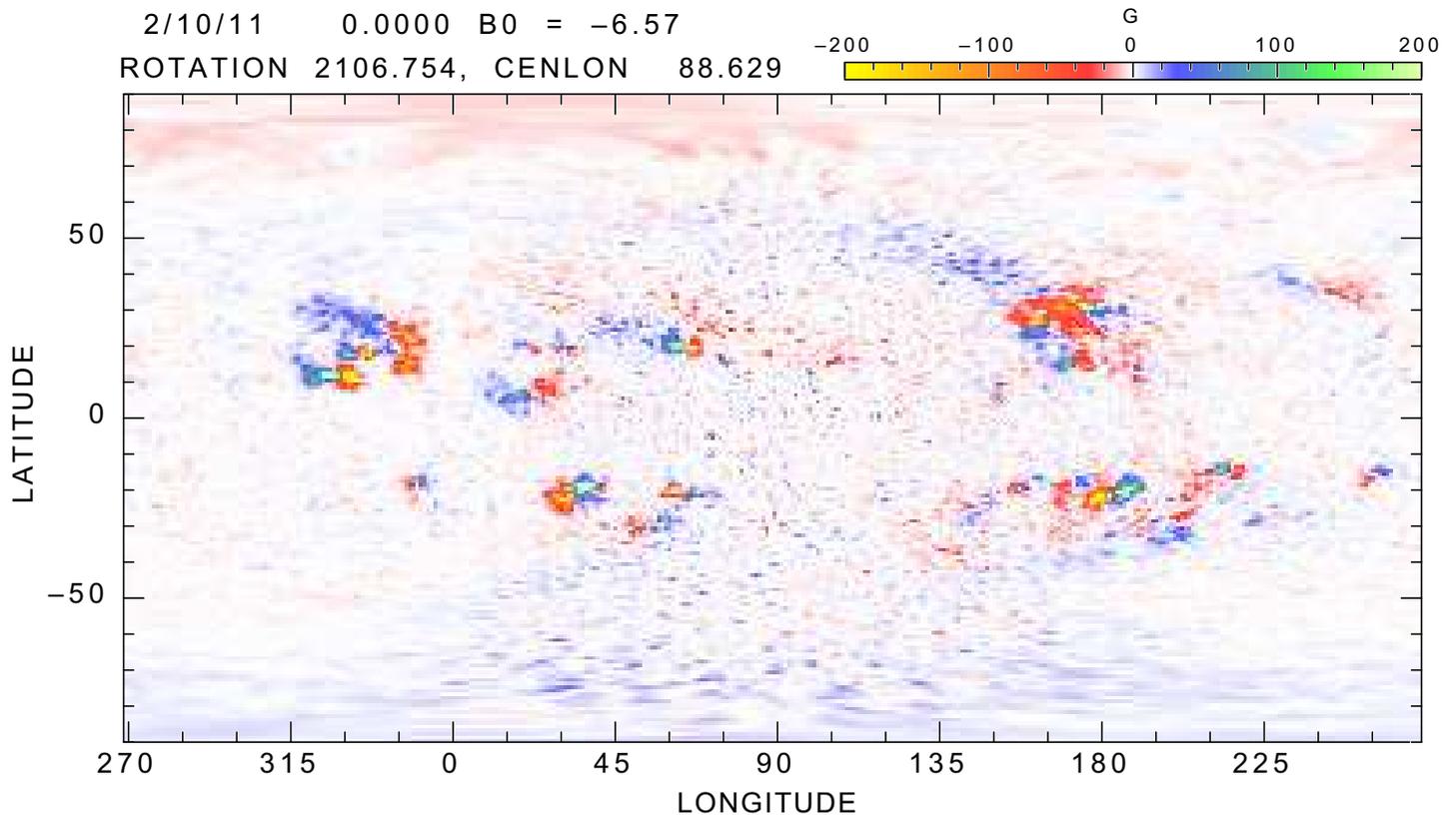}}
\caption{This figure gives a sample hybrid map.  The center part with a roughly hour-glass shape is from the HMI observations for the time of the map.  The quantity shown is the radial
component of the field vector after removal of the $B_\phi$ component.  Near the limb, especially in the polar regions,
the conversion from line-of-sight field to radial field produces a large enhancement of the resulting field strength.  The video 
at: {\tt http://www.astro.ucla.edu/\~{}ulrich/btss/Bro.mp4}
shows this result.  
In the north there are some frames for which the
matrix inversion of the time-dependent analysis \citep{2006SoPh...235...17U} has failed so that the non-time dependent values have been used.}
\label{figtwo}
}
\end{center}
\end{figure}

\begin{figure}\begin{center}
\hspace*{-0.2in}\parbox{7.0in}{\hspace*{0.3in}
\resizebox{6.5in}{!}{\includegraphics{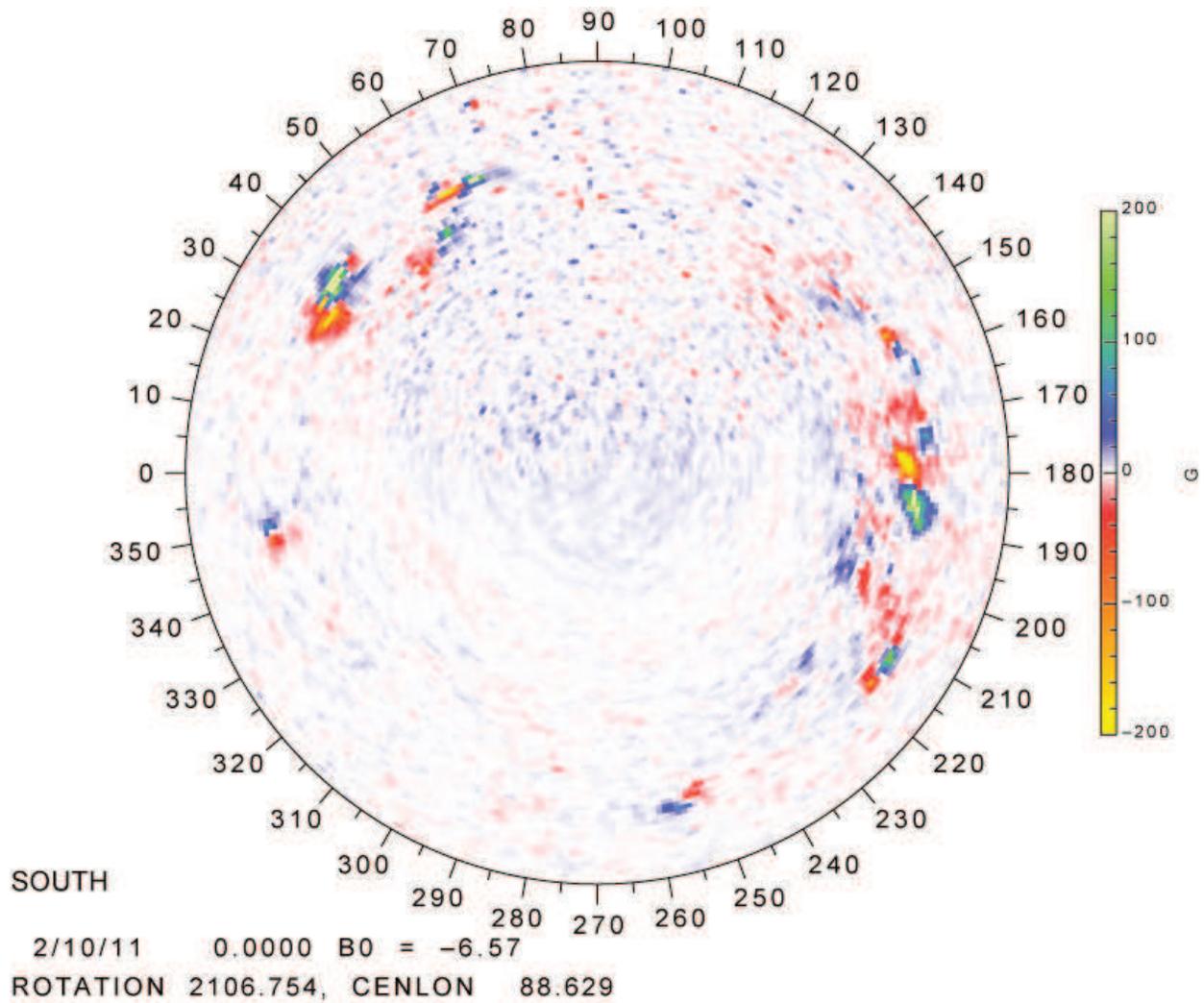}}
\caption{\baselineskip=15pt
This figure shows a remapping into a view as if from above the south pole for the map of Figure~\ref{figtwo}. The
quantity shown is the radial component of the photospheric magnetic field $B_r$.  Consequently this is not a simulation
of a line-of-sight magnetogram as might be seen from the vantage point along the axis of rotation looking toward the
south pole.  For regions near the polar axis it is close to such a magnetogram but for points progressively nearer
the outer edge of the circle the line-of-sight projection factor is missing and the values are much larger than would be seen.
The advantage of this map is the representation of the structures that make up the magnetic carpet.  The video from which
this frame is taken is 
at:
{\tt http://www.astro.ucla.edu/\~{}ulrich/btss/{\tt BroS.mp4}}.
Another advantage of
this mapping is that the solar coordinates are stationary while the position of the observer moves around the map making the variations of the field to be distinguished from the translation due to solar rotation which is part of the video 
presentation from Figure~\ref{figtwo}.}
\label{figthree}
}
\end{center}
\end{figure}

We show in Figure~\ref{figtwo} the hybrid map where the observed HMI fields have been interpolated into the snapshot map where the values replace the long-term estimated values coming from the UB analysis. The full sample time period is shown in the supplementary data video 
{\tt Bro.mp4}.  For all these maps the observed central meridian is shown at the center of the snapshot map and the Carrington longitude is given as the abscissa label.  In this representation, the solar magnetic fields flow through the map along with the short term variability induced by the supergranulation.  As an additional method of showing the magnetic fields in the polar zone we have taken latitude strip and wrapped them into circles having the radius they project as viewed along the polar axis but leaving
the fields on the map at the same value they have on the snapshot map.  These maps then show the geometry of the field as if it were being studied by an observer a large distance away from the solar south pole.  A sample of this plot is shown in Figure~\ref{figthree} and the supplementary data video is 
{\tt BroS.mp4}.  This set of map projections allows the inspection of the temporal variations of the flux concentrations in the important polar areas.  The longitudes are held fixed  while the region with current data fills a region that moves around the map.  The buffeting, disappearance and emergence of the flux concentrations is evident in this video.

\begin{figure}\begin{center}
\parbox{7.0in}{
\resizebox{7.0in}{!}{\includegraphics{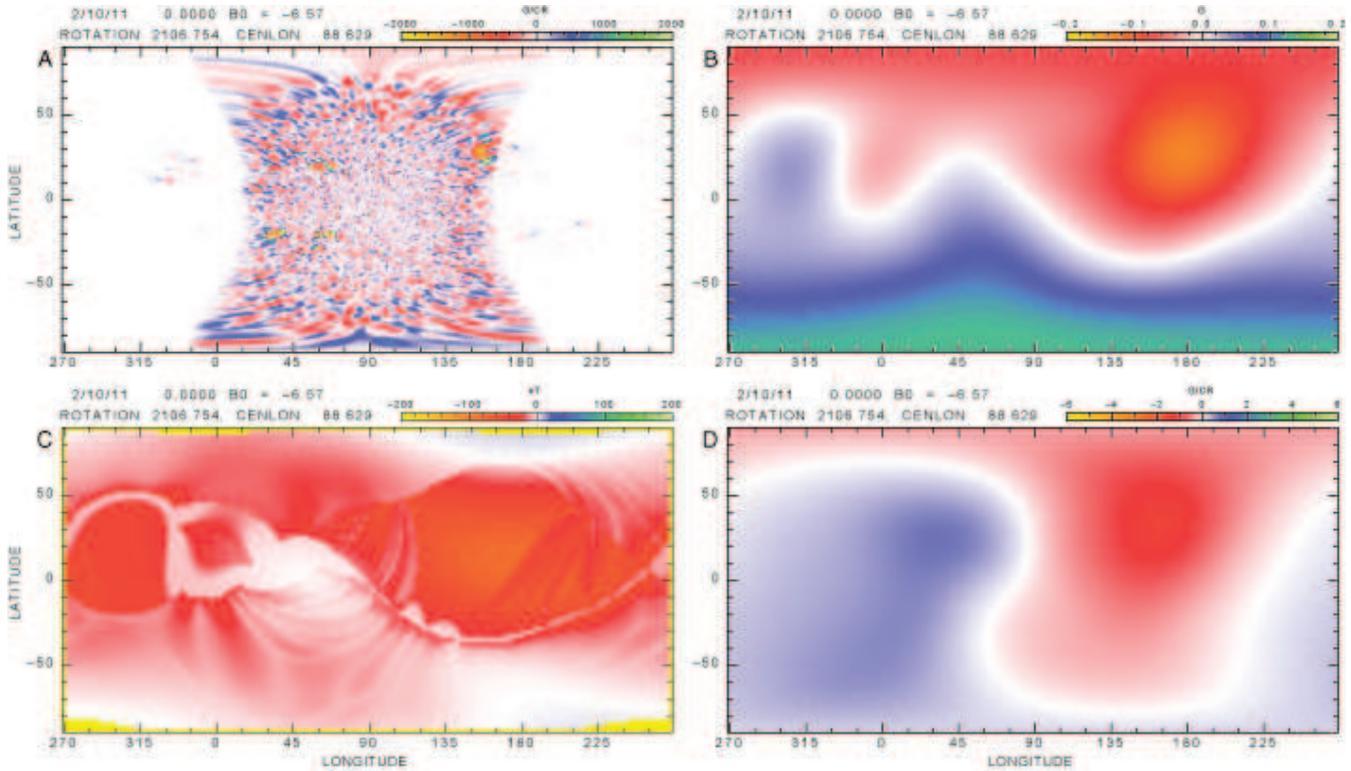}}
\caption{\baselineskip=13.5pt
This figure shows three steps toward the computation of the map of $B_\theta$ along with the resulting map.  The video 
{\tt http://www.astro.ucla.edu/\~{}ulrich/btss/FourPanel.mp4}
shows these same four maps. Panel A gives the photospheric map of the time derivative of the radial magnetic field $\dot B_r$ with the time units
being a Carrington rotation of 27.2753 days.  The rapid variations associated with the magnetic carpet are only available for those portions of the solar surface which can be observed and are much larger than the derivatives calculated from the long-time trends.  For the direct difference calculation we have adopted as our primary method, the time derivatives of panel A do not enter the reduction but we have retained this figure to illustrate where the variations originate.  Panel B gives the map of $B_r$ at the source surface as computed from the potential field source 
surface model. Such maps are normally shown with time separations of days or more.  When they viewed as part of the video where successive frames are separated by 2 h, the continuous fluctuation of the fields is a new feature.  The time derivative
of the radial magnetic field at the source surface $(\dot B_r)_{\rm ss}$ shown in panel D emphasizes the strength of this variability.  The unit of the time variation is one Carrington rotation.  Finally panel C shows the resulting map of $B_\theta$.  Because the calculation depends on the estimate of the solar wind speed at the source surface which includes a dependence on the flux tube expansion factor, some areas have a geometry which introduces a bounded region near the location where the field goes through zero. The overall direction of the $B_\theta$ field is governed by the fluctuations of the field in the polar regions so that, for example, when the field in the north is trending more negative (red) while the south is trending more positive (blue), the $B_\theta$ result trends more negative (red).}
\label{figthreea}
}
\end{center}
\end{figure}

The above plots and videos show that the overall level of the $B_r$ values is consistent between the UB reduction and the $B_\phi$ corrected los observations from HMI.  Because of the variability of the magnetic carpet structures, the values of $\dot B_r$ calculated from first differences between $(B_r)_{i+1}-(B_r)_{i-1}$ provide a rather different result where the amplitudes from the corrected HMI data are larger than that from the UB reduction by an amount that is roughly the ratio of the time steps, {\it i.e}. 27.27 days/2 h.  
The extension of the photospheric time dependent behavior out to the source surface is dominated by the largest scale averages of the structures.  Although there is high amplitude time dependence from the active latitude band, it has a relatively small spatial scale while the polar regions with a smaller amplitude exert a large influence the global structure of the source surface, especially for the dipole component.  The polar field variation introduces what amounts to a wobble in the dipolar structure with the consequent migration of the fields in the intermediate latitudes and this migration produces the $B_\theta$ component.  Figure~\ref{figthreea} along with the supplementary data video
{\tt FourPanel.mp4} show how this works to influence $(B_\theta)_{\rm ss}$.  The four maps all refer to the same time at each frame of the video from which the maps of Figure~\ref{figthreea} have been extracted.  The starting point is the photospheric time derivative map shown in the upper left (Figure~\ref{figthreea}A) which emphasizes the amplitude disparity between the current data and the long-term data as discussed above.  The fact that the maps use latitude as the ordinate means that the variable polar features appear outsized compared to the equatorial features.  The video associated with Figure~\ref{figthree} shows that this is an artifact of our map shape.  Figure~\ref{figthreea}B shows a familiar map of the radial component of the magnetic field at the source surface.  The upper right panel of the associated video demonstrates that in fact the source surface map undergoes a continuous jitter from the polar magnetic carpet variations.  The time derivative of the radial field at the source surface, $(\dot B_r)_{\rm ss}$, in Figure~\ref{figthreea}D shows some relationship to the $(B_r)_{\rm ss}$ map but is generally quite different in its structure.  We used two methods of calculating $(\dot B_r)_{\rm ss}$: a potential field expansion of the photospheric $\dot B_r$ and a direct difference of potential field expansions at two separated times.  We prefer the direct differences method because the potential field expansion of $B_r$ was less sensitive to individual wild points.    Equation~(\ref{Bthetaeqn}) requires an estimate of the solar wind speed.  For this we used a fit of solar wind speed $v_{\rm wind}$ to the square root of the flux tube expansion factor $f_{\rm s}^{1/2}$ using values given in Table 1 of \citet{1990ApJ...355..726W} yielding the numerical formula: $v_{\rm wind} = 835 - 105\;f_{\rm s}^{1/2} + 3.4\;f_{\rm s}\ {\rm km\;s}^{-1}$.  We have bounded the resulting wind speed to be greater than 200 km s$^{-1}$ and less than 800 km s$^{-1}$.  This approach is simplistic at best but is adequate for our exploratory work.  The solutions for $(B_\theta)_{\rm ss}$ are given in Figure~\ref{figthreea}C.  Of particular importance for the transverse field is the relationship between the north and south polar time derivatives.  At times where the polar time derivatives have the opposite sign, the global transverse field maps of $(B_\theta)_{\rm ss}$ tend to have a globally uniform sign reflecting the northward or southward drift of the field lines caused by this dipole field wobble.

\section{A Sample Time Series and Power Spectral Analysis}
\label{sample}

\begin{figure}\begin{center}
\parbox{6.0in}{
\resizebox{6.0in}{!}{\includegraphics{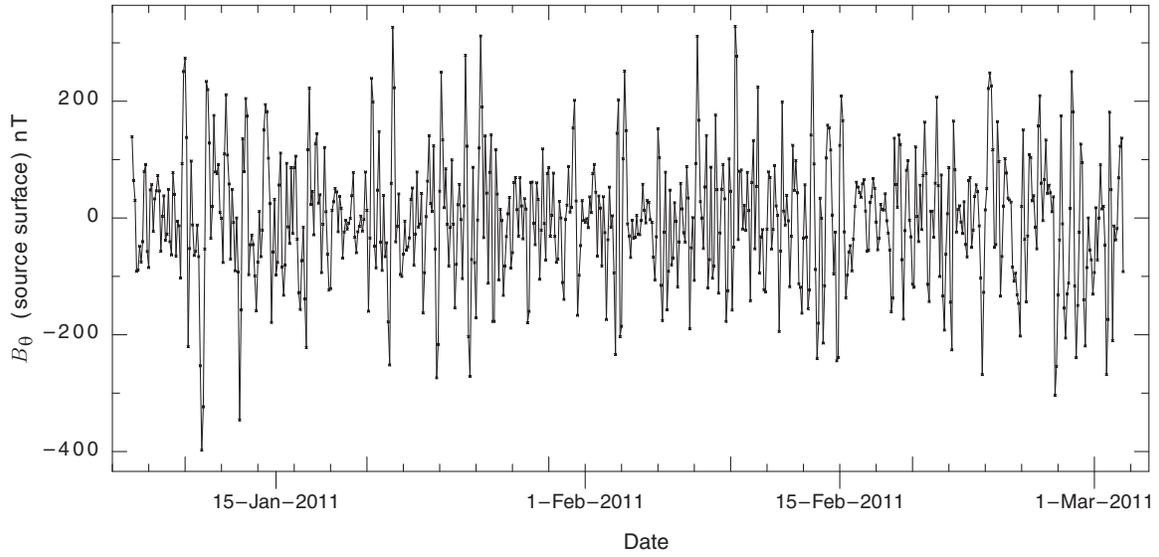}}
\caption{This figure shows the time dependence from the 2-h series of the average of $B_\theta$ on the source surface where the average is over a square 
$10^\circ\times10^\circ$ centered at a point at the latitude $b_0$ and at a longitude $20^\circ$ east of the central 
meridian.  The scale of the $y-$axis is at the source surface in nT and should be divided by approximately a factor 
of about 90 to apply to a point near Earth.  Possible effects due to the travel between the Sun and earth other than
this approximate strength decrease have not been included.}
\label{figfour}
}
\end{center}
\end{figure}

To illustrate the time series of $(B_\theta)_{\rm ss}$ quantitively we show in Figure~\ref{figfour} the average of this
quantity from the 2-h series over a square $10^\circ$ on a side centered $20^\circ$ east of the central meridian and at a latitude which
is on the ecliptic ({\it i.e.} the sub earth point). For these plots, we prepared a series which did not use the temporal filtering to avoid influencing the power spectrum.  This series is also based on time derivatives at the source surface from direct differences between pairs of PFSS solutions.   If the $B_\theta$ value from either series falls off from the source surface to the distance of 1 AU with a power law of $(r/2.5R_\solar)^{1.0}$ \citep{1989GeoRL..16....1J}, then the scale decrease would be about a factor of 90.  It is important
to note that we have not applied any adjustments to the calculation, including the suggested factor of about 1.4 derived by \citet{2012SoPh..279..295L} from comparison between HMI and MDI, and have omitted processes in the interplanetary region that could concentrate the fields due to compression as faster wind overtakes slower moving plasma.
 The amplitude of $B_{\rm z}$ at 1 AU is of order 1 to a few nT so a factor of
90 decrease in $B_\theta$ between the source surface and the 1 AU distance will bring the model results into reasonable
agreement with observations.  However, without an explicit transport model carrying the field from the source surface to a point of observation and a direct comparison to a set of observations, this agreement can say nothing about uncertainties such as the possible rescaling of the photospheric field strength observations due to uncertainties like those discussed by \citet{2012SoPh..279..295L}. 

\begin{figure}\begin{center}
\parbox{6.0in}{
\resizebox{6.0in}{!}{\includegraphics{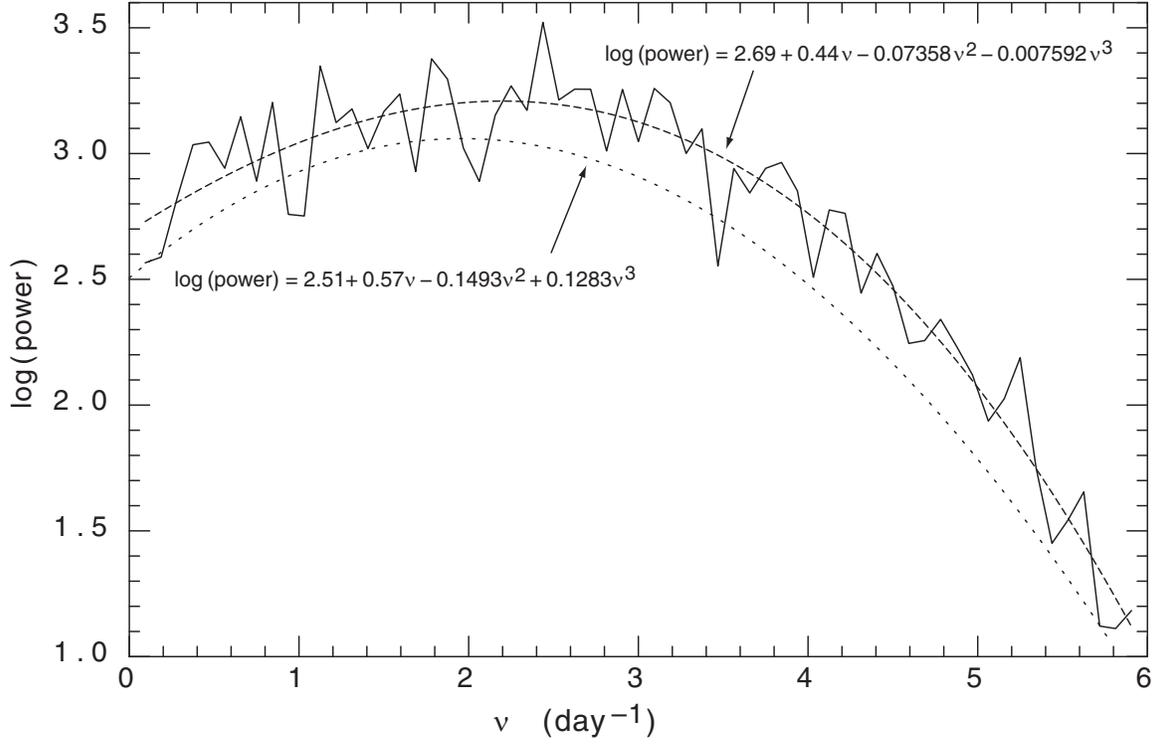}}
\caption{This figure shows power spectrum of $B_\theta$ derived from the time series shown in Figure~\ref{figfour}.  That series includes 655 points.  A power spectrum of the full series shows no long-term features apart from that at 1 day$^{-1}$ which we removed using a superposed epoch method.  The above spectrum is the average of the spectra from nine subsets of 128 measurements each.  The subset start times were separated by 64 points so that the time periods overlap and the resulting spectrum is oversampled.  The dashed line is a fit of a third order polynomial to the log of the spectrum.  This fit has a maximum at 2.2 day$^{-1}$ or about 11 h.  The upper formula gives the fit to this average spectrum in a numerical form.  The lower short-dashed line shows the fit to the power spectrum from the case where the polar contribution was taken from the slowly varying 2CRtd data set instead of from the {\sf Bro} data set.  Its fitting coefficients are shown below and to the left.}
\label{figfive}
}
\end{center}
\end{figure}

The power spectrum of the full 655 point data set revealed only a feature at one day which is likely an artifact from the HMI system.  We cleaned this artifact by subtracting from the data a superposed epoch average with a period of one day.  Then to obtain a smooth result, we prepared nine subsets of the data each having 128 points and with the starting point of each separated from the next by 64 points so that the subsets overlap.  This oversampling approach yields nine power spectra that we averaged together.  The power spectrum of $(B_\theta)_{\rm ss}$ of this average is given in Figure~\ref{figfive} in a log(power) {\it versus} frequency format.  The Nyquist frequency for this reduction is at 6 day$^{-1}$.  Note the very strong fall-off of the power at the
higher frequencies.  Prior to adopting the 2-h series as our primary data set, we had studied a 6-h series extensively and although the amplitude of the variation in $(B_\theta)_{\rm ss}$ increases strongly between the 6-h series and the 2-h series, the rapid falloff of the higher frequency spectrum indicates that there should not be a continued increase in the amplitude in the case of a possible still shorter sampling interval.  In addition, the peak in spectral density at about 11-h is roughly consistent with an interpretation that it is supergranular buffeting of high latitude magnetic fields that leads to the erratic changes in the global field configuration. 

\section{The Influence of High Latitudes and Polar Filling}

As a test of the importance of the high latitude contribution to the variability of $B_\theta$, we took values of
the polar region fields from the slowly varying 2CRtd map portion of the hybrid maps and treated the resulting field configuration the same way as the best estimate solution.  Specifically we replaced the {\sf Bro} values with the 2CRtd values using an interpolation that went from unity (100\%\ {\sf Bro} values) at $|{\rm latitude}|=60^\circ$ to zero (100\%\ 2CRtd values) at $|{\rm latitude}|=70^\circ$.  Our hybrid maps are like Mercator projections with a dimension of 360$\times$180.  As described above in Section~\ref{methods} the transition from the {\sf Bro} data to the 2CRtd data uses a weighting algorithm which depends on both center-to-limb angle and central meridian angle.  The effective area for which we use {\sf Bro} is the sum of the pixel weights times an additional factor of $\cos({\rm latitude})$ due to the spherical coordinate geometry.  The result of this test was that the rms variation in the test $10^\circ \times 10^\circ$ squares discussed in Section~\ref{sample} above decreased by 21\%\ while the decrease in effective area was only 7\%. This strong impact of the polar regions comes from the fact that the high latitudes couple well with the dipole moment and the variation in this component is responsible for the globally coherent behavior of $B_\theta$.  The fit of the power spectrum of the series without the polar areas is shown in Figure~\ref{figfive} as the short dashed line and the fitting coefficients are given below the line.  The overall shape remains the same with a decrease in power due to the lower amplitude as noted here also with a slight shift in shape so that the intermediate-to-higher frequency power is decreased a bit more than the lower frequency range.  This probably indicates a noise contribution from the near-limb pixels but could also indicate a change in the physical mechanism governing the convective motions in the poles compared to the more active regions.

In addition, the north pole regions were not visible during the selected sample period and the field values
were obtained from the \citet{2009PhDT.........1T} method.    To evaluate the influence these filled values might have on the solutions, we carried out another reduction where we did not carry out the  fill algorithm on the {\sf Bro} images to obtain current field values for the unseen portions of the solar surface.  In a calculation like that above we determined that the effective area of the {\sf Bro} data was reduced by 1.3\%\ while the rms variation in the derived $B_\theta$ was reduced by 9.9\%.

\section{Discussion and Conclusions}

We have developed and implemented a method of calculating $B_\theta$, the component of the interplanetary magnetic field which lies in the Sun's N/S plane.  We applied this method to a sample
period of two Carrington rotations in early 2011 and determined that the predicted amplitude of variations for $B_\theta$ is comparable to what is observed at 1AU.  Supergranular motions at high latitudes cause the $B_\theta$ variations during most periods due to the transport, emergence, and disappearance of moderate scale magnetic structures referred to as a magnetic carpet.  When we first 
encountered the large time derivatives in the photosphere, we were convinced they were an error of
some sort.  Perhaps it should not have been a surprise to us since \citet{2007ARA&A..45..297Z} has warned: ``There is a second process of field emergence that occurs
ubiquitously over the entire solar surface at much smaller spatial scales of less than
20,000 km. This small-scale process occurs during all parts of the solar cycle and, with
regard to the total emerging flux, actually dominates active regions."  We were able to overcome
our initial doubts by tracking individual points on the solar surface using the appropriate local rotation rate and found that in fact the variations were reproducible on a pointwise basis.  It is important to note further that the polar regions have a larger-than-average influence on $B_\theta$ variations.

A major shortcoming of our treatment is the lack of current data from the unseen portions of the solar surface.  These must have similar amplitude compared to what is available from HMI and will influence the global magnetic structure.  Hopefully the near side fields have a larger effect than the far side fields.
Within the conceptual model that the variations come from this convective phenomenon, there is little chance that the value of $B_\theta$ or $B_{\rm z}$ can be predicted without input from up-to-date observations having a high cadence like that from the HMI system.  The impact of larger events like coronal mass ejections (CMEs) and flares cannot be determined from our short sample but at least one CME appears to have introduced a larger than average excursion in $B_\theta$.
 An important ingredient in the calculation is the use of $\dot B_r$, the time derivative of the magnetic field component in the radial direction $B_r$.  This quantity has not been studied previously on a global scale and its calculation is challenging.  The high cadence of magnetic field measurements available from the HMI system has enabled our investigation.  

The present analysis is far short of what is needed to provide a prediction of $B_{\rm z}$ at 1 AU or elsewhere in the heliosphere.  We have relied on the very simple approach of the potential-field source-surface algorithm and a simple estimation of the solar wind speed as a function of a flux tube expansion factor.  Tracing the values of $B_\theta$ from the source surface at 2.5$R_\solar$ to the vicinity of interplanetary spacecraft which can measure $B_{\rm z}$ requires careful analysis that we have not carried out. 

Finally we note that although we have emphasized application to the $\theta$-component of $\mbf B$, the results for the $B_\phi$ component are similar.  It is quite likely that the helicity generated from the variations of both $B_\theta$ and $B_\phi$ together will be non-zero so that torsional Alfv\'en waves could be produced.   

\acknowledgments
We thank Jack Harvey for several helpful suggestions including bringing our attention to the perceptive quote from Zurbuchen.  We thank Chen Shi for uncovering an error in the sign of $B_\theta$ in our original equations.  This error is corrected in the current paper.
The HMI images used in preparation of this paper are courtesy of NASA/SDO and the science team.  This research has been supported by NASA through award NNX15AF39G to Predictive Science, Inc. and subaward to UCLA.

\bibliographystyle{spr-mp-sola}
\bibliography{rku-lib-1}
\newpage

\end{document}